\documentclass[aps,prl,twocolumn,groupedaddress]{revtex4-2}
\usepackage{amsmath,amsthm,amssymb,mathtools,tikz,booktabs,tensor,slashed,dutchcal,upgreek}
\usepackage[mathscr]{eucal}
\usetikzlibrary{cd}

\newcommand{\mc}{\mathcal} \newcommand{\mf}{\mathfrak} \newcommand{\mb}{\mathbb} \newcommand{\on}{\operatorname}  \newcommand{\ms}{\mathscr}
 \newcommand{\la}{\langle} \newcommand{\ra}{\rangle}
 \newcommand{\di}{\slashed D} \newcommand{\tr}{\on{Tr}}

\DeclareMathOperator{\gr}{\mathcal{R}}
\newcommand{\ccm}{\text{ccm}}

\allowdisplaybreaks

\begin{document}

\title{Kodaira--Spencer theory for flux backgrounds}



\author{Caleb Jonker}
\email{caleb.jonker@matfyz.cuni.cz}
\affiliation{Mathematical Institute, Faculty of Mathematics and Physics, Charles University, Prague 186 75, Czech Republic}
\author{Julian Kupka}
\email{j.kupka@herts.ac.uk}
\author{Ingmar Saberi}
\email{i.a.saberi@herts.ac.uk}
\author{Charles Strickland-Constable}
\email{c.strickland-constable@herts.ac.uk}
\affiliation{Department of Physics, Astronomy and Mathematics,
University of Hertfordshire, College Lane, Hatfield, AL10 9AB, United Kingdom}
\author{Fridrich Valach}
\email{fridrich.valach@matfyz.cuni.cz}
\affiliation{Mathematical Institute, Faculty of Mathematics and Physics, Charles University, Prague 186 75, Czech Republic}


\date{\today}

\begin{abstract}
We give an explicit description (in component fields) of a holomorphic theory associated to a general supersymmetric background of $\ms N=1$ supergravity in ten dimensions. Conjecturally, this provides a sought-for holomorphic realisation of the supergravity twist in such backgrounds, generalising the minimal type I BCOV theory for Calabi--Yau manifolds. Our theory unpacks the recently introduced Courant contact model associated to a holomorphic Courant algebroid. We also show that a coisotropic reduction of this model reproduces the recent model of ref~\cite{Ashmore:2025fxr}, which is formulated in terms of constrained fields.
\end{abstract}

\maketitle
\section{Introduction}
Ten dimensional $\ms N=1$ supergravity coupled to super Yang--Mills theory, first constructed in \cite{Bergshoeff:1981um,Chapline:1982ww,Dine:1985rz}, plays a prominent role in string theory, arising in particular in the low-energy limit of type I and heterotic string theory. Recently, the framework of generalised geometry has been used to explicitly construct its full Batalin--Vilkovisky (BV) description \cite{bvshort,bvlong}.

In \cite{Costello:2016mgj} Costello and Li introduced the notion of twist of any given supergravity as the formal expansion of the associated BV action around a critical point with nonvanishing supersymmetry ghost. We will call such a configuration a \emph{twisted background}. In the simplest possible case with all the fluxes vanishing, such a background corresponds to a Calabi--Yau metric and a choice of a covariantly constant pure spinor. Costello and Li conjectured that in this case the twists of ten-dimensional supergravities can be identified with particular flavours of the associated BCOV (or Kodaira--Spencer) theory \cite{Bershadsky:1993cx}.

The recent BV formulation of the $\ms N=1$ theory in particular provides an opportunity for checking this conjecture by a direct calculation. Unfortunately, although explicit formulas exist on both sides of the correspondence, the checking of their equivalence is still highly nontrivial and open at present. Nevertheless, the generalised-geometric formulation bears fruit already at this stage, as it suggests what the twist of the theory should be for an arbitrary twisted background --- including nonzero $H$-flux, curved gauge connection, and a nontrivial dilaton.

More specifically, as described in \cite{ccm}, in the context of $\ms N=1$, $D=10$ supergravity there is a natural generalised-geometric construction
\[\text{twisted background} \xmapsto{\text{Courant contact model}} \text{BV theory}\]
which, when applied to a Calabi--Yau case, outputs (the cotangent description of) the corresponding Kodaira--Spencer theory \cite{Costello:2019jsy,Raghavendran:2021qbh}. The aim of the present letter is to describe explicitly the output of this construction starting with a general background. The result can be seen below in formulas \eqref{f} and \eqref{oa}.

In the last section we show that after a suitable coisotropic reduction of the Courant contact model one recovers the recently introduced model \cite{Ashmore:2025fxr}, which was formulated in terms of constrained fields. The latter model was constructed in an independent fashion, as the ten-dimensional analogue of the six-dimensional superpotential functional. Correspondingly, its purely bosonic critical points involve Maurer--Cartan elements for the deformation problem of (F-term) supersymmetric heterotic backgrounds. The gauge sector of this model becomes holomorphic Chern--Simons theory, matching the holomorphic twist of super Yang--Mills theory that should appear upon coupling to supergravity in a twisted background. Further, on Calabi--Yau backgrounds its anomaly polynomial matches the physical supergravity and in the case without vector multiplets it was shown to match type I BCOV theory.
For these reasons, it was conjectured to describe the twist of $\ms N=1$ supergravity on a generic supersymmetric background. 
Our result here implies that the Courant contact model for a general twisted background provides a version of this model in which the constrained field space is resolved into a space of smooth unconstrained fields, as is needed for a proper BV treatment~\cite{CG2}. 
This implies the consistency of the conjecture of~\cite{Ashmore:2025fxr} with that of \cite{ccm}, which says:
\begin{center}
  \emph{The twist of $\ms N=1$ supergravity in ten dimensions is the Courant contact model for the twisted background.}
\end{center}

\section{Generalised geometry}
Fixing a principal $G$-bundle $P$ (for $G$ compact) the bosonic field content of $\ms N=1$, $D=10$ supergravity is given by a metric $g$, 3-form $H$, dilaton $\varphi$, and a $G$-connection $\nabla$, subject to the constraint \footnote{We have absorbed the usual coefficient on the RHS into the quadratic form $\text{Tr}$. Note that the low-energy effective description of type I string theory contains an additional $R^2$ term, necessary for the anomaly cancellation.}
\begin{equation}
  dH=\tr(F\wedge F),
\end{equation}
  where $F$ is the curvature of $\nabla$.
  Following \cite{Coimbra:2014qaa, direct}, we now describe how this data can be encoded more abstractly, using the language of generalised geometry, in terms of a triplet
  \[(\text{Courant algebroid},\;\text{generalised metric},\;\text{half-density}).\]
  Concretely, the Courant algebroid consists of the vector bundle (here $\on{ad}(P)$ is the adjoint bundle)
  \begin{equation}
    E=TM\oplus \on{ad}(P)\oplus T^*M
  \end{equation}
  equipped with the following bracket on its sections:
  \begin{align}\label{bracket}
    [x+s+\alpha&,y+t+\beta]=L_xy+(\nabla_xt-\nabla_ys+[s,t]_\mf g\nonumber\\
    &+i_yi_xF)+(L_x\beta-i_yd\alpha+i_yi_xH\\
    &+\tr[(\nabla s)t-(i_xF)t+(i_yF)s]).\nonumber
  \end{align}
  It also carries a pairing
  \begin{equation}\label{pairing}
    \la x+s+\alpha,y+t+\beta\ra=\alpha(y)+\beta(x)+\tr(st)
  \end{equation}
  and the obvious projection map (called the \emph{anchor})
  \begin{equation}\label{anchor}
    \rho\colon E\to TM,\qquad \rho(x+s+\alpha)=x,
  \end{equation}
  and all this data satisfies a suitable set of axioms \cite{liu1997manin,let}.
  
  A generalised metric is by definition the datum of an orthogonal splitting $E=C_+\oplus C_-$. In this case we take the splitting to be determined by
  \begin{equation}
    C_+=\on{graph}(g)\equiv\{x+g(x,\cdot)\mid x\in TM\}.
  \end{equation}
  Finally, the half-density (on $M$) is $\sigma=e^{-\varphi}\sqrt[4]{|g|}$.
  
  One of the advantages of the generalised-geometric framework is the fact that various operations and constructions take a very succinct and conceptually clear form. For instance, the entire bosonic part of the supergravity action becomes simply
  \begin{equation}
    S_{\text{bosonic}}=\int_M \gr\sigma^2,
  \end{equation}
  where $\gr$ is the generalised scalar curvature, constructed analogously to the ordinary scalar curvature by treating the Courant algebroid (with its bracket) as the analogue of the tangent bundle (with the commutator of vector fields). Similarly, the fermionic part of the action \cite{direct, full} simplifies drastically, as do the supersymmetry transformations. In particular, this led to the recent construction of the BV action for the supergravity in \cite{bvshort,bvlong}.
  
  The implication for the present work is the characterisation of the twisted background, which now depends on the generalised metric, half-density, and the supersymmetry ghost $e$. The latter is a positive chirality spinor associated to the subbundle $C_+$, twisted by (i.e.\ with values in) half-densities on $M$. The twisted background conditions, describing the criticality of the BV action, take the elegant form
  \begin{equation}\label{twist}
    \bar e \gamma_a e=0,\qquad \di e=0,\qquad D_\alpha e=0.
  \end{equation}
  Here the Latin and Greek indices refer to the subbundles $C_+$ and $C_-$, respectively, $D$ is a \emph{generalised Levi-Civita connection}, and $\di=\gamma^a D_a$. It follows (see e.g\ \cite{ccm}) that
  \begin{equation}
    L:=\on{annihilator}(e)\subset C_+\otimes \mb C
  \end{equation}
  is a rank 10 isotropic subbundle of $E\otimes\mb C$ whose sections close under the bracket \eqref{bracket}. This in particular defines a complex structure on $M$, with $T^{0,1}M=\rho(L)$.
  
  \section{Courant contact model}
  Before focusing on the details of the concrete model, let us briefly sketch the abstract construction of the Courant contact model \cite{ccm} associated to the above twisted background.
  
  First, the properties of $L$ imply that the bracket, pairing, and anchor of $E$ descend to the corresponding operations on $L$-invariant sections of $L^\perp/L$, turning the latter object into a \emph{holomorphic Courant algebroid} \cite{GenKaehler}. This can then be resolved to yield a complex
  \begin{equation}
    (\mc E^\bullet=\Omega^{0,\bullet}(L^\perp/L),\bar\partial_\mc E)
  \end{equation}
  of vector bundles, equipped again with a bracket $[\cdot,\cdot]_\mc E$, pairing $\la \cdot,\cdot\ra_\mc E$, and an anchor map $\rho_\mc E$ \footnote{The bracket with sections of $L$ descends to a flat $L$-connection on the complex smooth vector bundle $L^\perp/L$. Since $L\cong T^{0,1}M$ this is equivalent to a Dolbeault operator on $\Omega^{0,\bullet}(L^\perp/L)$, which we have denoted $\bar\partial_\mc E$.}. The Courant contact model associated to the twisted background is then the theory with the field space
  \begin{equation}\label{fccm}
    \ms F_{\ccm}:=\Omega^{0,\bullet}\times \Pi\mc E^\bullet\times (\Omega^{5,\bullet})^\times,
  \end{equation}
  where $\Pi$ denotes the parity reversal \footnote{Note that due to the lack of R-symmetry which could be used for regrading, the twist of the $\ms N=1$ theory in ten dimensions is expected to only carry a $\mb Z_2$-grading (instead of a full $\mb Z$-grading).}, and $(\Omega^{5,\bullet})^\times$ denotes $\Omega^{5,\bullet}$-forms with everywhere nonzero $\Omega^{5,0}$-part.
  Let us denote a general element of $\ms F_{\ccm}$ by $(f,\xi,\lambda)$.
  The field space carries an odd BV symplectic form
  \begin{equation}\label{occm}
    \omega_{\ccm}=\delta\int_M\lambda(\delta f+\tfrac12\la \xi,\delta\xi\ra_\mc E)
  \end{equation}
  as well as an (even) action functional
  \begin{equation}\label{accm}
    S_{\ccm}=\int_M\lambda(\rho_\mc E(\xi)f+\tfrac16\la \xi,[\xi,\xi]_\mc E\ra_\mc E+\bar\partial f+\tfrac12\la \xi,\bar\partial_\mc E\xi\ra_\mc E)
  \end{equation}
  which satisfies the \emph{classical master equation}
  \begin{equation}
    \{S_{\ccm},S_{\ccm}\}=0.
  \end{equation}
  The conjecture put forward in \cite{ccm} is that this model coincides with the twist of supergravity for a chosen twisted background.
  
\section{Decomposing the construction}
  Let us now unpack this construction in terms of the original supergravity variables. We start with the twist data conditions. Regarding $\zeta:=\sigma^{-1}e$ as an ordinary (tangent space) spinor w.r.t.\ $g$, equations \eqref{twist} can be rewritten as (cf.\ Appendix C.4 of \cite{direct})
  \begin{equation}
    \begin{aligned}
      \nabla_\mu \zeta-\tfrac18H_{\mu\nu\rho}\gamma^{\nu\rho}\zeta&=0,\\
      \bar\zeta\gamma_\mu\zeta=\tfrac1{2}\slashed H\zeta-(\nabla_\mu\varphi)\gamma^\mu \zeta=\slashed F\zeta&=0.
    \end{aligned}
  \end{equation}
  In particular $\zeta$ is the pure spinor associated to the aforementioned complex structure $I$ on $M$. Setting $\omega:=gI$, the equations in particular imply
  \begin{equation}
    H=-d^c\omega,\quad i_\omega H=d^c\varphi,\quad i_\omega F=0,\quad F^{0,2}=0,
  \end{equation}
  where $d^c=i(\bar\partial-\partial)$.
  The last condition ensures that the original smooth principal bundle $P$ with connection $\nabla$ gives rise to a holomorphic principal bundle $P_\mb C$ with connection $\nabla^\mb C$ w.r.t.\ the complexified Lie group $G_\mb C$ \footnote{More explicitly, the natural map $G\to G_\mb C$ allows us to construct $P_\mb C$ as the associated bundle $P\times_G G_\mb C$. This (still regarded as a smooth object) inherits a connection, denoted $\nabla^\mb C$. In particular the tangent space at each point in $P^\mb C$ can be identified with the direct sum of a tangent space on the base with the Lie algebra $\mf g_\mb C$. Since both of these spaces carry a complex structure, we may attempt to combine them into a complex structure on $P^\mb C$. This indeed works --- the integrability is ensured by the condition $F^{0,2}=0$.}.
  
  The reduced holomorphic Courant algebroid was described in \cite{Garcia-Fernandez:2020awc}: its resolved version $\mc E$ takes the form
  \begin{equation}
    \mc E^\bullet=\Omega^{0,\bullet}(T^{1,0}M\oplus \on{ad}(P_\mb C)\oplus T^{*1,0}M).
  \end{equation}
  The remaining operations are
  \begin{equation}
    \bar \partial_\mc E=\begin{pmatrix}
    \bar\partial & 0 & 0\\
    F^{1,1}_\mb C&(\nabla^\mb C)^{0,1}&0\\
    2i\partial \omega& -F^{1,1}_\mb C&\bar\partial
  \end{pmatrix},
  \end{equation}
  \begin{align}
      [x+s+\alpha&,y+t+\beta]_\mc E=L^\partial_xy+(\nabla^\mb C_xt-\nabla^\mb C_ys+[s,t]_{\mf g^\mb C})\nonumber\\
      &+(L^\partial_x\beta-i_y\partial\alpha+\tr (\nabla^\mb C_{1,0} s)t),
  \end{align}
  where $L^\partial$ is the holomorphic Lie derivative, extended naturally to the Dolbeault-resolved complex (cf.\ Example 3.12 of \cite{ccm}). The pairing and anchor map are formally the same as \eqref{pairing} and \eqref{anchor}.
  
  The resulting Courant contact model then has the field space $\ms F_{\ccm}$ given by
  \begin{align}\label{f}
    \Omega^{0,\bullet}\times\! \Pi\Omega^{0,\bullet}(T^{1,0}M\!\oplus \on{ad}(P_\mb C)\!\oplus T^{*1,0}M)\!\times\! (\Omega^{5,\bullet})^\times
  \end{align}
  with fields denoted $f,x,s,\alpha,\lambda$. The symplectic form and action are
  \begin{align}\label{oa}
    \omega_{\ccm}&=\delta\int_M\lambda(\delta f+\tfrac12i_x\delta\alpha+\tfrac12i_{\delta x}\alpha+\tfrac12\tr(s\delta s))\nonumber\\
    S_{\ccm}&=\int_M\lambda (L_x^\partial f+\tfrac12i_x L_x^\partial\alpha+\tfrac12\tr(s\nabla^\mb C_xs)\\
    &\quad\vphantom{\int_M}+\tfrac16\tr (s[s,s])+\bar\partial f+\tfrac12i_x \bar\partial\alpha+\tfrac12i_{\bar\partial x}\alpha\nonumber\\
    &\quad\vphantom{\int_M}+\tfrac12\tr (s\nabla^\mb C_{0,1}s)-\tr (si_x F^{1,1}_\mb C)-i_xi_x (i\partial\omega)).\nonumber
  \end{align}
  
  We note that, as appropriate for a twisted theory, the resulting model only depends on the holomorphic part of the original physical data. This is conveniently encoded in the intermediate holomorphic Courant algebroid structure.
  
  It is also interesting to study critical points of this theory; the simplest are the ones with $f_0,x_0,s_0,\alpha_0$ vanishing and $\lambda_0=\Omega\in\Omega^{5,0}$ invertible. The criticality condition is then $\bar\partial\Omega=0$. Thus the existence of such a critical point forces $M$ to be a Calabi--Yau 5-fold \footnote{Here Calabi--Yau manifold is understood as a complex manifold with a holomorphic volume form.}. Expanding the differential $Q=\{S_{\ccm},\cdot\}$ around $\Omega$ we then obtain (cf.\ formula (10) in \cite{ccm})
  \begin{align}
    Q&=\int_M\left(\bar\partial f+\tfrac12L_x^\partial f-\tfrac14i_x L_x^\partial\alpha-\tfrac14\tr(s\nabla^\mb C_xs)\right.\nonumber\\
    &\qquad-\left.\tfrac1{12}\tr (s[s,s]_{\mf g^\mb C})\right)\frac{\delta}{\delta f}+(\bar\partial x+\tfrac12L_x^\partial x)\frac{\delta}{\delta x}\nonumber\\
    &\qquad+(\nabla^\mb C_{0,1}s-i_x F^{1,1}_\mb C+\nabla^\mb C_xs+\tfrac12[s,s]_{\mf g^\mb C})\frac{\delta}{\delta s}\\
    &\qquad+(\bar\partial\alpha-2i_x(i \partial\omega)-\tr (F^{1,1}_\mb Cs)+\partial f+\tfrac12L_x^\partial \alpha\vphantom{\frac{\delta}{\delta \alpha}}\nonumber\\
    &\qquad+\tfrac12i_x\partial\alpha+\tfrac12\tr s \nabla^\mb C_{1,0} s)\frac{\delta}{\delta \alpha}+(\bar\partial\lambda+L_x^\partial\lambda) \frac{\delta}{\delta \lambda}.\nonumber
  \end{align}
  As usual, the linear, quadratic and cubic terms encode the unary, binary, and ternary brackets of the associated $L_\infty$ algebra.
  
\section{Reducing to the constrained model}
  Fixing now a holomorphic volume form $\Omega$ as above, let us show how the Courant contact model recovers the constrained theory of \cite{Ashmore:2025fxr} via a coisotropic reduction. We will in fact not need the explicit formulas \eqref{f}--\eqref{oa} but can work already on the level of the general construction \eqref{fccm}--\eqref{accm}.
  
  The field space of \cite{Ashmore:2025fxr} is given by the middle-degree coholomogy of the complex
  \begin{equation}
    \ms C_\Omega:=\; \Omega^{0,\bullet}\xrightarrow{\mc D}\Pi\mc E^\bullet\xrightarrow{\on{div}_\Omega}\Omega^{0,\bullet},
  \end{equation}
  where $\mc D$ is defined by $\la u,\mc Df\ra_\mc E=\rho_\mc E(\xi)f$. We note that this complex has an invariant pairing in which the norm of $(f,\xi,\vartheta)$ is
  \begin{equation}
    \int_M\Omega(f\wedge\vartheta+\tfrac12\la \xi,\xi\ra),
  \end{equation}
  and with respect to which we have $\mc D^*=\text{div}_\Omega$. 
  
  In order to recover this model from the Courant contact model we start by considering the subspace
  \begin{equation}
    \ms S_\Omega=\{(f,\xi,\lambda)\mid \lambda=\Omega,\;\on{div}_\Omega\xi=0\}\subset \ms F_{\ccm},
  \end{equation}
  where $\on{div}_\Omega\xi:=\Omega^{-1}L_{\rho_\mc E(\xi)}^\partial \Omega$. For the tangent space at any $p\in \ms S_\Omega$ we then get $T_p\ms F_{\ccm}\cong \Omega^{0,\bullet}\oplus\Pi\mc E^\bullet\oplus \Omega^{5,\bullet}$, and
  \begin{equation}
    T_p\ms S_\Omega\cong \Omega^{0,\bullet}\oplus\Pi(\ker(\text{div}_\Omega))\oplus 0.
  \end{equation}
  
  Note that the natural embedding $T_p\ms S_\Omega\to \ms C_\Omega$ intertwines $\omega_{\ccm}|_{\ms S_\Omega}$ with the pairing on $\ms C_\Omega$. Finally, calculating the symplectic orthogonal of $T_p\ms S_\Omega$ w.r.t.\ $\omega_{\ccm}$ we obtain
  \begin{equation}
    (T_p\ms S_\Omega)^{\perp}=\Omega^{0,\bullet}\oplus \Pi(\ker(\text{div}_\Omega))^\perp\oplus 0,
  \end{equation}
  where on the RHS the $\perp$ is taken w.r.t.\ the pairing on $\ms C_\Omega$.
  
  Assuming $M$ is compact, we can now use Hodge theory to show \footnote{Using the explicit expressions for $\mc E^\bullet$ and the identification $\Omega^{0,\bullet}(T^{1,0}M)\cong \Omega^{4,\bullet}$ provided by the volume form, the equality $(\ker(\text{div}_\Omega))^\perp=\on{im}\mc D$ reduces to $[\ker(\partial\colon \Omega^{4,\bullet}\to\Omega^{5,\bullet})]^{\perp}=\partial\Omega^{0,\bullet}$, where $\perp$ on the LHS is taken w.r.t.\ the wedge-and-integrate pairing. Choosing now any auxiliary Hermitian metric, the LHS becomes $[\ker(\partial^\dagger\colon \Omega^{1,\bullet}\to\Omega^{0,\bullet})]^\perp$, with $\perp$ taken w.r.t.\ the metric. Using the Hodge decomposition we then conclude that $(\ker\partial^\dagger)^\perp=\on{im}\partial$.} that
  \begin{equation}
    (\ker(\text{div}_\Omega))^\perp=\on{im}\mc D,
  \end{equation}
  implying $(T_p\ms S_\Omega)^{\perp}\subset T_p\ms S_\Omega$, i.e.\ the subspace $\ms S_\Omega$ is coisotropic. We can therefore perform a coisotropic reduction by quotienting out the directions in $\ms S_\Omega$ in which $\omega_{\ccm}|_{\ms S_\Omega}$ is zero. The resulting reduced space $\ms F_{\text{red}}$ then precisely gives
  \begin{equation}
    \ms F_{\text{red}}=\frac{\{\xi\in\Pi\mc E^\bullet\mid \on{div}_\Omega\xi=0\}}{\{\mc Df\mid f\in\Omega^{0,\bullet}\}}.
  \end{equation}
  
  Similarly, restricting the action we get
  \begin{equation}
    S_{\ccm}|_{\ms S_\Omega}=\int_M\Omega(\tfrac12\la \xi,\bar\partial_\mc E\xi\ra_\mc E+\tfrac16\la \xi,[\xi,\xi]_\mc E\ra_\mc E).
  \end{equation}
  Since this is constant along the null directions, it descends to the quotient $\ms F_{\text{red}}$, where it defines an action functional $S_{\text{red}}$, which by construction satisfies the classical master equation. The result is precisely the constrained model of \cite{Ashmore:2025fxr}.
  
  The above procedure can be regarded as a particular case of (the classical variant of) the BV fibre integration procedure \cite{severa2004noncommutative,Cattaneo:2015vsa}, in which one restricts the path integrand to a coisotropic submanifold of the field space and then performs the path integral along the null directions of the BV symplectic form.
  We also remark that the relation between the model \cite{Ashmore:2025fxr} and the Courant contact model is akin to the relation between the Kodaira--Spencer theory \cite{Bershadsky:1993cx} (written in terms of fields satisfying differential constraints) and Costello and Li's later model, (minimal) BCOV theory \cite{Costello:2015xsa} (which uses a resolved description in terms of smooth fields).
  
\section{Conclusions}
  Starting from a twisted background we have produced a BV theory \eqref{f}--\eqref{oa}. The construction used the language of generalised geometry, with the result depending only on the holomorphic field data (encoded in the associated holomorphic Courant algebroid). In \cite{ccm} it was shown that in the fluxless Calabi--Yau case this model coincides with the corresponding BCOV theory, which is the conjectural description of the twist of $\ms N=1$ supergravity in 10 dimensions for such a background \cite{Costello:2019jsy,Raghavendran:2021qbh}. 
  
  It is thus natural to conjecture that the Courant contact model \eqref{f}--\eqref{oa} is the twisted theory associated to an arbitrary background with fluxes. 
It follows from the last section that this conjecture is consistent with that of~\cite{Ashmore:2025fxr}, as coisotropic reduction of our model reproduces the one found therein. However, our model \eqref{f}--\eqref{oa} has the advantage that it is written in terms of smooth fields, which resolve the constrained fields of~\cite{Ashmore:2025fxr}.
As another check, we note that the gauge part of the action \eqref{oa} correctly reproduces the holomorphic twist of the super Yang--Mills theory, i.e.\ the 10-dimensional holomorphic Chern--Simons theory.

\begin{acknowledgments}
C.S.-C.\ is supported by an EPSRC New Investigator Award, grant number EP/X014959/1. C.J.\ and F.V.\ are supported by the Charles University grant PRIMUS/25/SCI/018. 
    This work is funded by the Deutsche Forschungsgemeinschaft (DFG, German Research Foundation) under Projektnummer 517493862 (Homologische Algebra der Supersymmetrie: Lokalit\"at, Unitarit\"at, Dualit\"at; I.A.S. is co-P.I.).
We thank P.~\v{S}evera for valuable discussions. No new data was collected or generated during the course of this research.
\end{acknowledgments}
\bibliography{citations}

\end{document}